\documentclass[twocolumn,showpacs,preprintnumbers,amsmath,amssymb]
{revtex4-1}

\usepackage{color}
\usepackage{graphicx}
\usepackage{dcolumn}
\usepackage{bm}

\begin{document}

\preprint{}

\title{Quark tensor charge and electric dipole moment within the Schwinger-Dyson formalism
}

\author{Nodoka~Yamanaka}
  \email{yamanaka@ruby.scphys.kyoto-u.ac.jp}
  \affiliation{Department of Physics, Graduate School of Science,
  Kyoto University, \\
  Kitashirakawa-oiwake, Sakyo, Kyoto 606-8502, Japan}
\author{Takahiro~M.~Doi}
  \affiliation{Department of Physics, Graduate School of Science,
  Kyoto University, \\
  Kitashirakawa-oiwake, Sakyo, Kyoto 606-8502, Japan}
\author{Shotaro~Imai}
  \affiliation{Department of Physics, Graduate School of Science,
  Kyoto University, \\
  Kitashirakawa-oiwake, Sakyo, Kyoto 606-8502, Japan}
\author{Hideo~Suganuma}
  \affiliation{Department of Physics, Graduate School of Science,
  Kyoto University, \\
  Kitashirakawa-oiwake, Sakyo, Kyoto 606-8502, Japan}

\date{\today}

\begin{abstract}

We calculate the tensor charge of the quark in the QCD-like theory in the Landau gauge using the Schwinger-Dyson formalism.
It is found that the dressed tensor charge of the quark is significantly suppressed against the bare quark contribution, and the result agrees qualitatively with the analyses in the collinear factorization approach and lattice QCD.
We also analyze the quark confinement effect with the phenomenological strong coupling given by Richardson, and find that this contribution is small.
We show that the suppression of the quark tensor charge is due to the superposition of the spin flip of the quark arising from the successive emission of gluons which dress the tensor vertex.
We also consider the relation between the quark and the nucleon electric dipole moments by combining with the simple constituent quark model.
\end{abstract}


\maketitle

\section{\label{sec:intro}Introduction}

The analysis of the nucleon parton structure plays an essential role in the fundamental study of the quantum chromodynamics.
The quark distribution of the nucleon in the leading twist is given by the momentum distribution $f_1$, the spin distribution $g_1$, and the transversity distribution $h_1$ functions of the quark, and has been studied in high-energy experiments.
The transversity distribution gives the spin distribution of the quark carrying the momentum fraction $x$ of the total momentum of the transversely polarized nucleon.
The total transversity of the quarks inside the nucleon is given by the quark tensor charge, defined by the relation
\begin{equation}
\delta q = \int_0^1 dx \left[ h_1 (x) - \bar h_1 (x) \right]
\, ,
\end{equation}
where $h_1 (x)$ and $\bar h_1 (x)$ are the transversity distribution of the quark and antiquark in the nucleon.
The quark transversity distribution has been the focus of many theoretical investigations \cite{transversityreview,transversitytheory,transversitymodel,extraction,lattice}.
In the nonrelativistic limit, the tensor charge is the spin of the particle.
In the nonrelativistic constituent quark model, which considers three massive quarks in the nucleon, the tensor charge of the quark in the proton is thus given by $\delta u = \frac{4}{3}$ ($u$ quark) and $\delta d = -\frac{1}{3}$ ($d$ quark) \cite{adler}.

The transverse polarization of the quark in nucleons can be extracted from experimental observables involving the simultaneous polarization of the beam and the target, such as the semi-inclusive deep inelastic electron-nucleon scattering or the polarized Drell-Yang process.
The single-spin asymmetries for semi-inclusive deep-inelastic scattering with pion production can probe the quark transversity, and were measured experimentally by the HERMES \cite{hermes} and COMPASS \cite{compass} collaborations.
Recently, the first extraction of the quark transversity distribution from these experimental data using the collinear factorization approach became available \cite{extraction}, and the total tensor charge (at the renormalization point $\mu =1$ GeV) was given by 
\begin{eqnarray}
\delta u &=& 0.57 \pm 0.21 \, , \nonumber\\
\delta d &=& -0.18 \pm 0.33 \, .
\label{eq:extraction}
\end{eqnarray}
Despite the large theoretical uncertainty, this result shows a suppression compared with the constituent quark model prediction.

Also the lattice QCD studies of the quark tensor charge have been done so far \cite{lattice}, and they also predict values suppressed against the constituent quark model prediction.
The typical result with the lattice QCD simulation (S. Aoki {\it et al.} in Ref. \cite{lattice}) is
\begin{eqnarray}
\delta u &=& 0.839 \pm 0.060 \, , \nonumber\\
\delta d &=& -0.231 \pm 0.055 \, , \nonumber\\
\delta s &=& -0.046 \pm 0.034 \, ,
\label{eq:latticeqcd}
\end{eqnarray}
where the renormalization point was fixed to $\mu \simeq 1.4$ GeV.
This suppression is consistent with the tensor charge extracted from the experimental data [Eq. (\ref{eq:extraction})]. 
It is now of importance to clarify the source of this suppression.

The importance of the investigation of the tensor charge is not restricted in the study of the nucleon structure function.
This quantity is actually useful in the analysis of the neutron electric dipole moment (EDM).
The neutron EDM is an observable sensitive to the $CP$ violation of the hadronic system and is thus an excellent probe of new physics beyond the standard model \cite{pospelovreview}.
The current experimental data of the neutron EDM are given by $d_n < 2.9 \times 10^{-26}e\, {\rm cm}$ \cite{baker}, which can provide many constraints on the $CP$ violation of the new physics such as the supersymmetric models \cite{pospelovreview,mssm}.
In many candidates of theories beyond the standard model, $CP$ violating interactions give a large contribution to the electric dipole moment of quarks.
In such situations, we need to know the dependence of the quark EDM on the neutron EDM.
Many works with this motivation exist in the literature \cite{hecht,degrassi,nagata,pertquarkmodel}.
The EDM of the neutron $d_n$ is defined by the limit of zero momentum transfer of the $CP$-odd nucleon form factor.
The dependence of the neutron EDM on the quark EDM $d_q$ is related to the tensor charge by \cite{bhattacharya}
\begin{equation}
d_n 
= \sum_q d_q \delta q 
\, .
\end{equation}
This means that the sensitivity of the neutron EDM on the new physics beyond the standard model depends on the tensor charge of the quarks, so whether the quark tensor charge is small or not thus becomes one of the main points of interest.

In watching the suppression of the quark tensor charge extracted from the experimental data or from the lattice QCD simulations against the constituent quark model prediction, we note two sources of suppression can na\"{i}vely be inferred.
The first source is the dressing of the bare quark tensor charge by gluons, and the second possibility is the spin-dependent bound state effect.
The first case was not discussed previously, and should be treated nonperturbatively to extract the physics.

As a powerful nonperturbative way to investigate the dynamics of the quantum field theory and, in particular, the low energy QCD, we have the Schwinger-Dyson (SD) formalism, and many studies such as the dynamical quark mass, the meson masses, etc, have been done so far \cite{higashijima,miransky,kaoki,robertsreview,alkoferreview,pinchtech,pi-k,emformfactor,quarkpropagator}.
The effect in question, the vertex gluon dressing, is also well within the applicability of the SD formalism.
In this paper, we will therefore try to clarify the physics involved in the vertex dressing by gluons and analyze the source of the suppression of the quark tensor charge. 

This paper is organized as follows.
In Section \ref{sec:setup}, we give the formulation of the SD formalism, the renormalization improved running couplings used in this work, and a brief explanation of the derivation of the dynamical quark mass.
In Section \ref{sec:sdeq}, we formulate the SD equation for the quark tensor charge and give the result of the calculation.
In Section \ref{sec:analysis}, we compare our result with the collinear factorization approach and lattice QCD results, analyze the effect of the gluon dressing to the tensor vertex, and give the dependence of the neutron EDM on the quark EDM.
The renormalization of the quark EDM will also be discussed there.
The final section is devoted to the summary.

\section{\label{sec:setup}Basic Formalism}

In this section, we present the detail of the QCD-like theory and the quark propagator used in this paper.
We assume the rainbow-ladder approximation in which the nonperturbative effect of the gluon is included by improving the momentum dependence of the quark-gluon vertex \cite{quark-gluon_vertex} by the one-loop level renormalization group.
This gives the replacement
\begin{equation}
\frac{g_s^2}{4\pi} Z_g (q^2) \gamma^\mu \times \Gamma^\nu (q , k ) 
\rightarrow
\alpha_s (q^2) \gamma^\mu \times \gamma^\nu
\, ,
\label{eq:rlapprox}
\end{equation}
where $Z_g (q^2)$ is the gluon dressing function, and $\Gamma^\nu (q,k) $ is the dressed quark-gluon vertex.
In this work, we use the Landau gauge which minimizes the unphysical momentum fluctuation of the gluons in the Euclidean space-time.
To compare and discuss the result obtained, we use three different renormalization group improved strong couplings: the QCD running coupling (one-loop level, $N_f =3$) with infrared (IR) regularization \`{a} la Higashijima \cite{higashijima}, the smooth IR regularization \cite{kaoki}, and the running coupling with the Landau pole shifted to zero momentum (Richardson Ansatz) \cite{richardson}.
We use the QCD scale parameter $\Lambda_{\rm QCD}$ = 900 MeV for the analysis without approximation, and $\Lambda_{\rm QCD}$ = 500 MeV when the Higashijima-Miransky approximation is used (the ordinary QCD scale parameter is around $\Lambda_{\rm QCD} \simeq 200 - 300$ MeV. In this paper, the large scale parameter is taken to reproduce the chiral quantities).

\begin{figure}[htb]
\includegraphics[width=6.1cm,angle=-90]{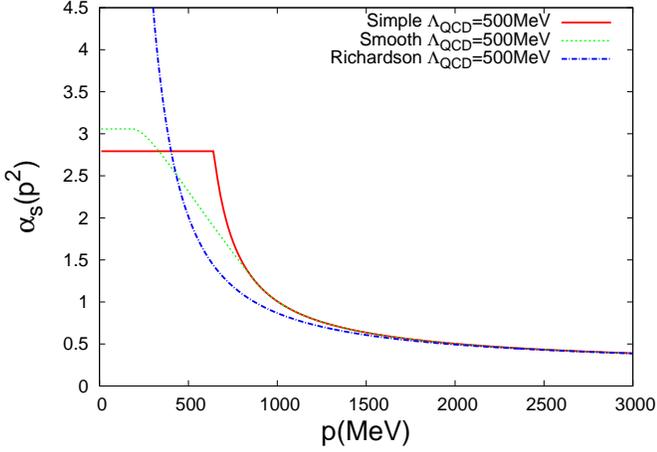}
\caption{\label{fig:strong_coupling}
The running strong couplings of QCD-like theory.
We use the running couplings with the simple infrared regularization, the smooth infrared regularization, and the Richardson Ansatz.
}
\end{figure}

The first running strong coupling with the simple IR regularization is defined by \cite{higashijima}
\begin{eqnarray}
\alpha_s (p^2) =
\left\{
\begin{array}{ll}
\frac{24\pi}{11N_c - 2N_f} & (p<p_{\rm IR}) \cr
\frac{12\pi}{11N_c - 2N_f} \frac{1}{\ln (p^2 / \Lambda_{\rm QCD}^2)} & (p\geq p_{\rm IR}) \cr
\end{array}
\right.
\, .
\end{eqnarray}
where $N_c =3 $, and $p_{\rm IR}$ satisfies $\ln (p_{\rm IR}^2 / \Lambda_{\rm QCD}^2)= \frac{1}{2}$.
As it can be seen in Fig. \ref{fig:strong_coupling}, this running coupling has one cusp in the infrared region.
This IR regularization was introduced to avoid the divergent Landau pole at $p=\Lambda_{\rm QCD}$.

The second running strong coupling with smooth IR regularization is defined by \cite{kaoki}
\begin{eqnarray}
\alpha_s (p^2) 
&=&
\frac{3C_2 (N_c)}{16\pi \beta_0} 
\nonumber\\
&&
\times
\left\{
\begin{array}{ll}
C & (p \leq p_0 ) \cr
C -\frac{1}{2} \frac{1}{\ln^2 \left( \frac{p_{\rm IR}^2}{ \Lambda_{\rm QCD}^2 } \right)} 
\frac{\ln^2 \left( p^2 / p_0^2 \right)}{\ln  \left( p_{\rm IR}^2 / p_0^2 \right)}
& (p_0 < p < p_{\rm IR}) \cr
\frac{1}{\ln (p^2 / \Lambda_{\rm QCD}^2)} & (p\geq p_{\rm IR}) \cr
\end{array}
\right.
\, .
\nonumber\\
\end{eqnarray}
where the lowest coefficient of the $\beta$ function of the renormalization group is given by $\beta_0 = \frac{11N_c -2N_f }{ 48\pi^2}$, and $C=\frac{1}{2} \frac{\ln  \left( p_{\rm IR}^2 / p_0^2 \right)}{\ln^2 \left( \frac{p_{\rm IR}^2}{ \Lambda_{\rm QCD}^2 } \right)} + \frac{1}{\ln \left( \frac{p_{\rm IR}^2}{ \Lambda_{\rm QCD}^2 } \right)} $.
Here we have set $\ln (p_{\rm IR}^2 / \Lambda_{\rm QCD}^2) = \frac{1}{2} $ and $\ln (p_0^2 / \Lambda_{\rm QCD}^2) = -2$. 
For this running coupling, the discontinuity of the derivative of the running coupling is removed, and we have no cusps in the IR region.

The running strong coupling with the Landau pole shifted to the zero momentum point $p=0$ (the Richardson Ansatz) is given by
\begin{equation}
\alpha_s (p^2)
=
\frac{12\pi}{11N_c - 2N_f} \frac{1}{\ln (1+p^2 / \Lambda_{\rm QCD}^2)} 
\, .
\end{equation}
This running coupling generates a linear confining potential $V(r) \simeq \sigma r - \frac{A}{r}$ in a phenomenological manner, where the string tension is given by $\sigma = \frac{C_2 ({\bf 3} ) \Lambda_{\rm QCD}^2}{8\pi \beta_0} $ and the Coulomb coefficient is given by $A=\frac{C_2 ({\bf 3} )}{8\pi \beta_0}$.
It is thus possible to analyze the effect of the quark confinement within this framework.
The string tension in this model is $\sigma \approx 1.2$ GeV/fm.
This value is slightly larger than the physical string tension $\sigma_{\rm phys} \approx 0.89$ GeV/fm.
In treating this running coupling numerically, we shift the pole by a very small number to avoid the divergence at $p=0$ MeV.
The shapes of the three running couplings are plotted in Fig. \ref{fig:strong_coupling}.

\begin{figure}[htb]
\begin{center}
\includegraphics[width=6.1cm,angle=-90]{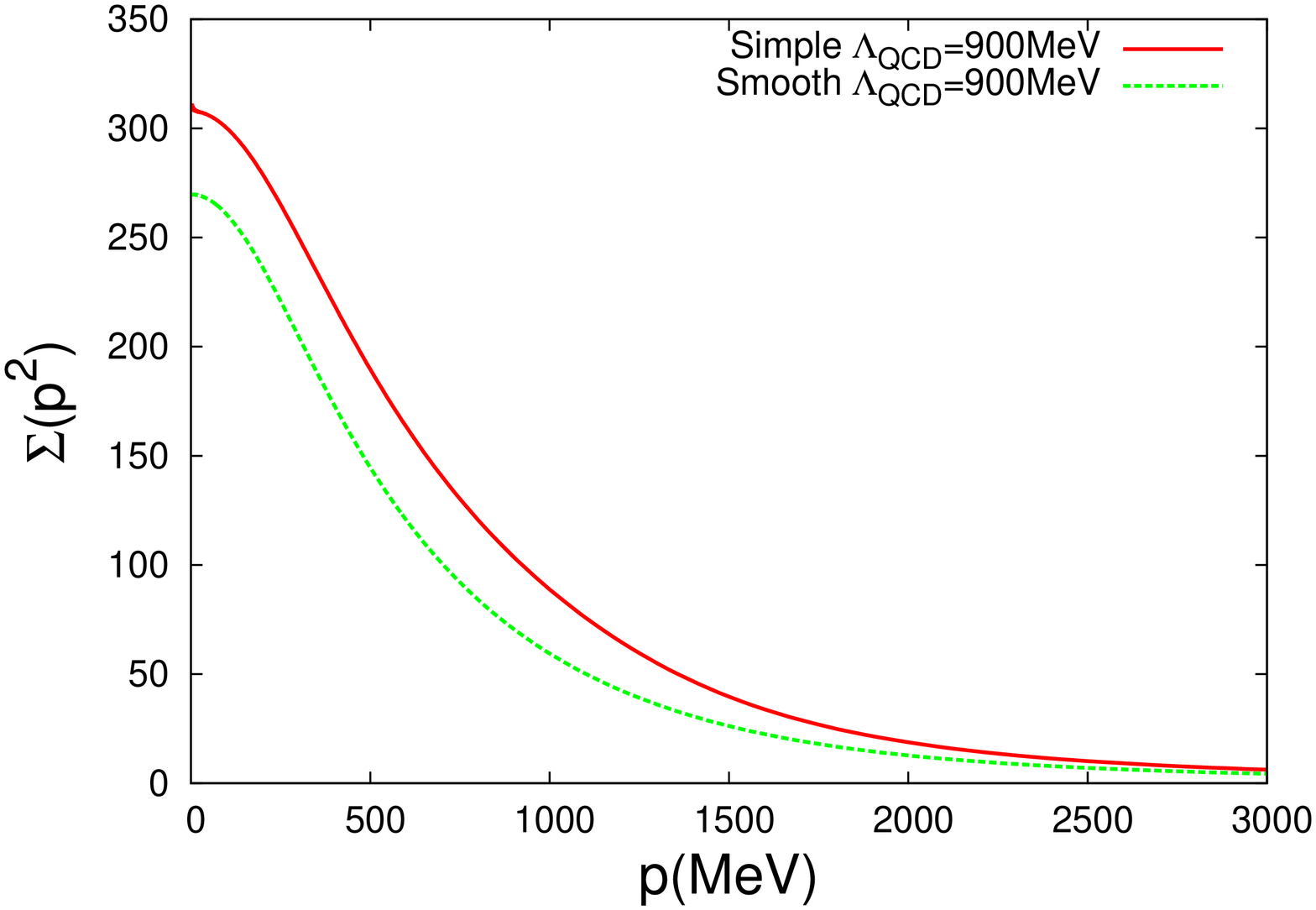}
\caption{\label{fig:self-energy_full}
The quark self-energy $\Sigma (p_E^2)$ solved with the Schwinger-Dyson equation.
}
\end{center}
\begin{center}
\includegraphics[width=6.1cm,angle=-90]{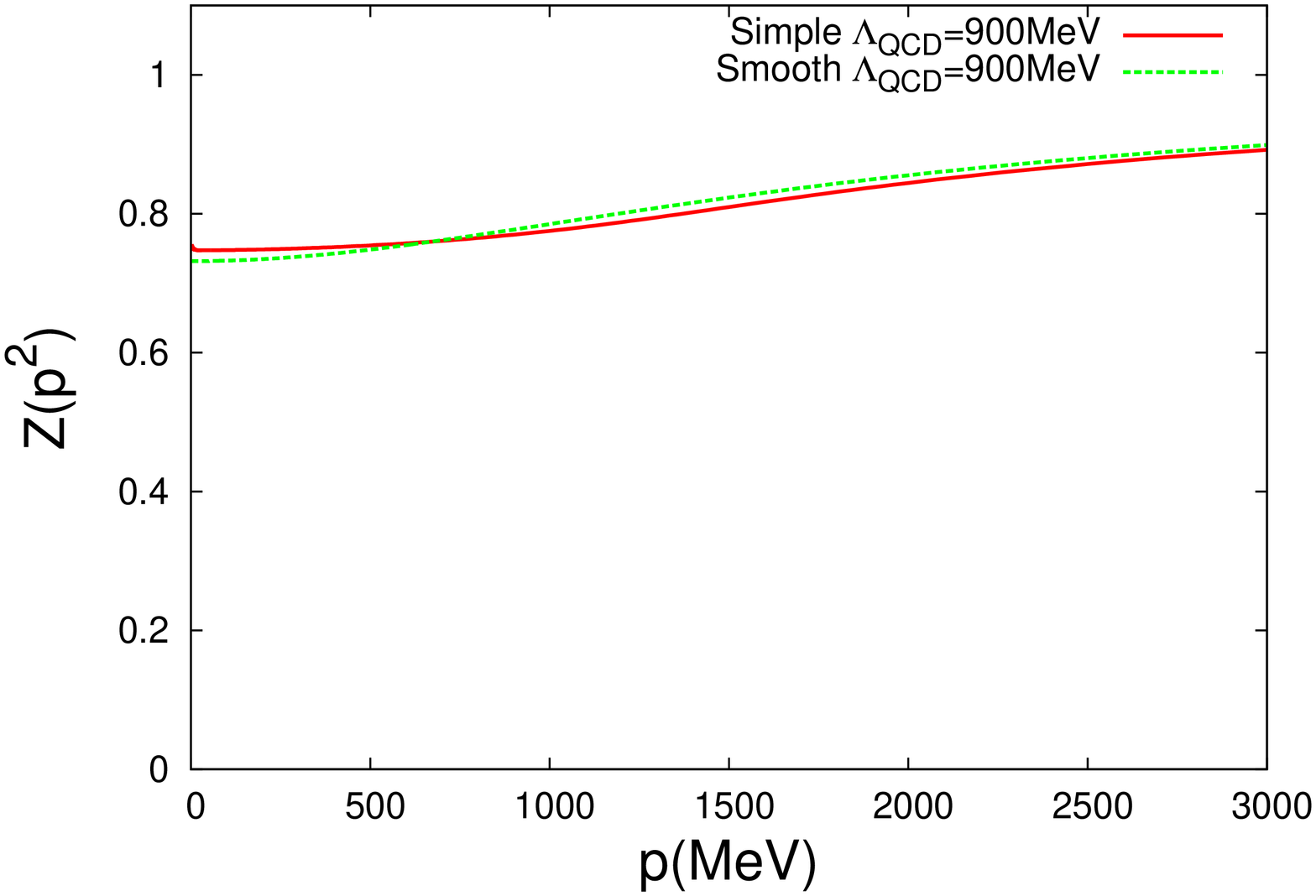}
\caption{\label{fig:wavfctnrenormalization_full}
The quark wave function renormalization $Z (p_E^2)$ solved with the Schwinger-Dyson equation.
}
\end{center}
\end{figure}

We now solve the quark propagator SD equation in the Landau gauge.
In this paper, we consider the SD equation with the effect of the dressed gluon propagator and dressed quark-gluon vertex included in the RG improved strong coupling [see Eq. (\ref{eq:rlapprox})].
The SD equation is a system of two integral equations
\begin{eqnarray}
\frac{\Sigma (p^2)}{Z(p^2)} 
&=&
m_q 
-\frac{3i \,C_2 (N_c)}{4\pi^3} \int \hspace{-.3em} d^4k
\frac{\alpha_s [(p-k)^2]}{(p-k)^2}
\nonumber\\
&&\hspace{10em} \times
\frac{Z(k^2) \Sigma (k^2)}{k^2 -\Sigma^2 (k^2)}
.
\label{eq:SD_eq4}
\\
\frac{1}{Z (p^2)}
&=&
1+i \frac{C_2 (N_c) }{8\pi^3 p^2} 
\int \hspace{-.3em} d^4k
\frac{ \alpha_s [(p-k)^2] }{k^2 -\Sigma^2 (k^2)}
Z(k^2)
\nonumber\\
&&\hspace{3em} \times
\left[
2-\frac{p^2+k^2}{(p-k)^2} -\frac{(p^2-k^2)^2}{(p-k)^4}
\right]
.
\label{eq:wavfctnsde}
\end{eqnarray}
where $Z(k^2)$ and $\Sigma (k^2)$ are the wave function renormalization and the self-energy of the quark, respectively.
In this paper, we take the chiral limit $m_q =0$.
The quark wave function renormalization and the quark self-energy are plotted in Figs. \ref{fig:self-energy_full} and \ref{fig:wavfctnrenormalization_full}, respectively.
We see that the self-energy is generated dynamically even in the chiral limit.

\begin{figure}[htb]
\begin{center}
\includegraphics[width=6.1cm,angle=-90]{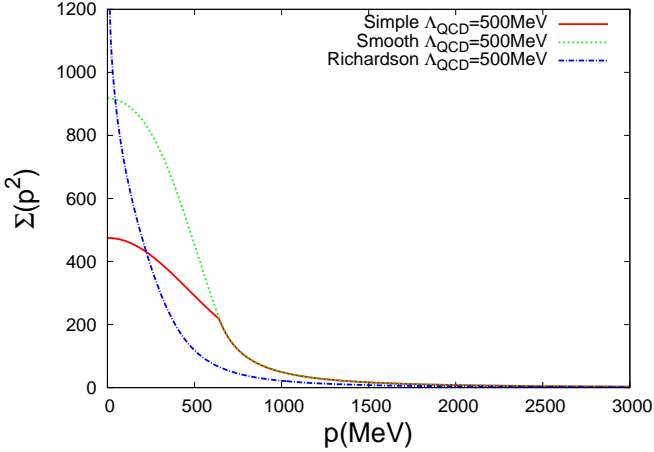}
\caption{\label{fig:self-energy}
The quark self-energy $\Sigma (p_E)$ solved with the Schwinger-Dyson equation with the Higashijima-Miransky approximation.
}
\end{center}
\end{figure}

Taking the Higashijima-Miransky approximation
\begin{equation}
\alpha_s [(p_E-k_E)^2]
\approx
\alpha_s \left[{\rm max} (p_E^2, k_E^2)\right]
\, ,
\label{eq:Higashijima-Miransky}
\end{equation}
with $p_E$ and $k_E$ the Euclidean momenta, we have 
\begin{eqnarray}
Z (p_E^2)
&=&
1
,
\\
\Sigma (p_E^2)
&=&
\frac{3C_2 (N_c)}{2\pi} \int_0^\Lambda
\frac{k_E^3 dk_E \Sigma (k_E^2)}{k_E^2 +\Sigma^2 (k_E^2)}
\frac{\alpha_s \left[ {\rm max} (p_E^2, k_E^2)\right]}{{\rm max} (p_E^2, k_E^2)}
. 
\nonumber\\
\end{eqnarray}
The resulting quark self-energy is plotted in Fig. \ref{fig:self-energy}.
We should note that the quark propagator SD equation is not calculable with the Richardson Ansatz, due to the singularity at $p-k =0$ (this forms a singular line in the phase space of $p^\mu$ and $k^\mu$).
In the Higashijima-Miransky approximation, however, this singularity is avoided by ${\rm max} (p_E^2, k_E^2)$, the only remaining singularity is the point $p^\mu=k^\mu=0$.
Numerically, this remaining singularity was avoided by shifting the pole by a small number, and we have verified that this shift does not change the resulting quark self-energy $\Sigma (p^2)$.
We can say that the Higashijima-Miransky approximation acts as a regularization in the Richardson Ansatz.

The quark self-energy can be related to the chiral condensate with
\begin{equation}
\langle \bar q q \rangle_\Lambda 
=
-
\frac{N_c }{ 2\pi^2} \int_0^{\Lambda} 
\hspace{-0.5em}
k_E^3 dk_E
\frac{Z (k_E^2) \Sigma(k_E^2)}{k_E^2 + \Sigma^2 (k_E^2)}
\, .
\label{eq:chiral_condensate}
\end{equation}
The parameter $\Lambda$ is the ultraviolet cutoff (not to be confused with $\Lambda_{\rm QCD}$).
In our numerical calculation, the cutoff was taken as $\Lambda =$ 20 GeV.
To  obtain the chiral condensate renormalized at $\mu =$ 2 GeV, we use the formula
\begin{equation}
\langle \bar q q \rangle_\mu
=
\left( \frac{\alpha_s (\Lambda^2 )}{\alpha_s (\mu^2 ) } \right)^{\frac{3C_2 (N_c)}{16\pi^2 \beta_0}}
\langle \bar q q \rangle_\Lambda
\, ,
\label{eq:renormalized_chiral_condensate}
\end{equation}
where $\frac{3C_2 (N_c)}{16\pi^2 \beta_0}=\frac{4}{9}$.
The above renormalized chiral condensate is stable in the variation of the cutoff scale $\Lambda$ [numerically, we have verified that the variation is small, of $O(10^{-3})$. See Tables \ref{table:stability} and \ref{table:stability_hm}].
This proves that the high-energy behavior of the quark propagator is well described in the SD formalism with the Higashijima-Miransky approximation.

From the quark self-energy, it is also possible to give the pion decay constant $f_\pi$ with the Pagels-Stokar approximation \cite{pagels}:
\begin{eqnarray}
f_\pi^2 
&=&
\frac{N_c }{2\pi^2} \int_0^\Lambda k_E^3 dk_E \, \frac{\Sigma (k_E^2 ) Z(k_E^2)}{\left[ k_E^2 +\Sigma^2 (k_E^2 ) \right]^2}
\nonumber\\
&&\hspace{6em} \times
\left[ \Sigma (k_E^2 ) - \frac{k_E}{4} \frac{d}{dk_E} \Sigma (k_E^2 ) \right]
. \ \ \ 
\label{eq:pagels-stokar}
\end{eqnarray}
The pion decay constant is an observable, so its renormalization is not required.
The chiral condensate and the pion decay constant obtained in this framework are shown in Table \ref{table:chiral_condensate}.
We have also calculated the same physical quantities in the Higashijima-Miransky approximation, which are given in Table \ref{table:chiral_condensate_hm}.

\begin{table}
\caption{The chiral condensate and the pion decay constant obtained from the self-energy calculated in the Schwinger-Dyson formalism.
The unit is in MeV [in (MeV)$^3$ for the chiral condensate].
The chiral condensate was calculated with Eq. (\ref{eq:renormalized_chiral_condensate}) at the renormalization point $\mu = 2$ GeV.
The pion decay constant was obtained from the Pagels-Stokar approximation (\ref{eq:pagels-stokar}) with the cutoff $\Lambda = 20$ GeV.
}
\begin{ruledtabular}
\begin{tabular}{ccccc}
IR behavior & $\Lambda_{\rm QCD}$& $\langle \bar q q \rangle_\mu$ & $f_\pi$ \\
\hline
Simple & 900 & $-(248)^3$ & 70  \\
Smooth & 900& $-(221)^3$ & 60 \\
\end{tabular}
\end{ruledtabular}
\label{table:chiral_condensate}
\end{table}

\begin{table}
\caption{The chiral condensate and the pion decay constant obtained from the self-energy calculated in the Schwinger-Dyson formalism with the Higashijima-Miransky approximation.
We have used the same parameters as Table \ref{table:chiral_condensate}.
}
\begin{ruledtabular}
\begin{tabular}{ccccc}
IR behavior & $\Lambda_{\rm QCD}$& $\langle \bar q q \rangle_\mu$ & $f_\pi$ \\
\hline
Simple & 500 & $-(242)^3$ & 90  \\
Smooth & 500 & $-(243)^3$ & 96 \\
Richardson & 500 & $-(193)^3$ & 66 \\
\end{tabular}
\end{ruledtabular}
\label{table:chiral_condensate_hm}
\end{table}

\section{\label{sec:sdeq}The Schwinger-Dyson equation for the quark tensor charge}

\begin{figure}[htb]
\includegraphics[width=8.2cm]{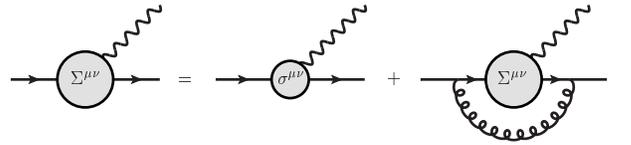}
\caption{\label{fig:tensor_SD_eq} 
The Schwinger-Dyson equation for the quark tensor charge expressed diagrammatically.
}
\end{figure}

Let us consider the SD equation of the quark tensor charge (or the quark EDM) depicted diagrammatically in Fig. \ref{fig:tensor_SD_eq}.
The SD equation for the quark tensor charge is given by
\begin{eqnarray}
\Sigma^{\mu \nu} (p)
&=&
\sigma^{\mu \nu}
\nonumber\\
&&
+
i C_2 (N_c) \int \frac{d^4k}{4\pi^3}
\alpha_s \left[(p-k)^2 \right] Z^2(k^2)
\nonumber\\
&& \hspace{1em} \times
\gamma^\rho \frac{k\hspace{-.45em}/\, + \Sigma (k^2)}{k^2 -\Sigma^2 (k^2)}
\Sigma^{\mu \nu} (k)
\frac{k\hspace{-.45em}/\, + \Sigma (k^2)}{k^2 -\Sigma^2 (k^2)} \gamma^\lambda 
\nonumber\\
&& \hspace{1em} \times 
D_{\rho \lambda} (p-k)
,
\label{eq:tensorsde}
\end{eqnarray}
where $ D_{\rho \lambda} (q) \equiv \frac{-1}{q^2} \left( g_{\rho \lambda} - \frac{q_\rho q_\lambda }{q^2} \right) $ is the gluon propagator in the Landau gauge (the color index was factorized), and $\Sigma^{\mu \nu}$ is the dynamical tensor charge in the zero limit of the momentum transfer.
As for the quark propagator SD equation, we consider the rainbow-ladder approximation [see Eq. (\ref{eq:rlapprox})] in which the effect of the dressed gluon propagator and the dressed quark-gluon vertex included in the renormalization group (RG) improved strong coupling given in the previous section.

\begin{figure*}[htb]
\includegraphics[width=17cm,angle=-90]{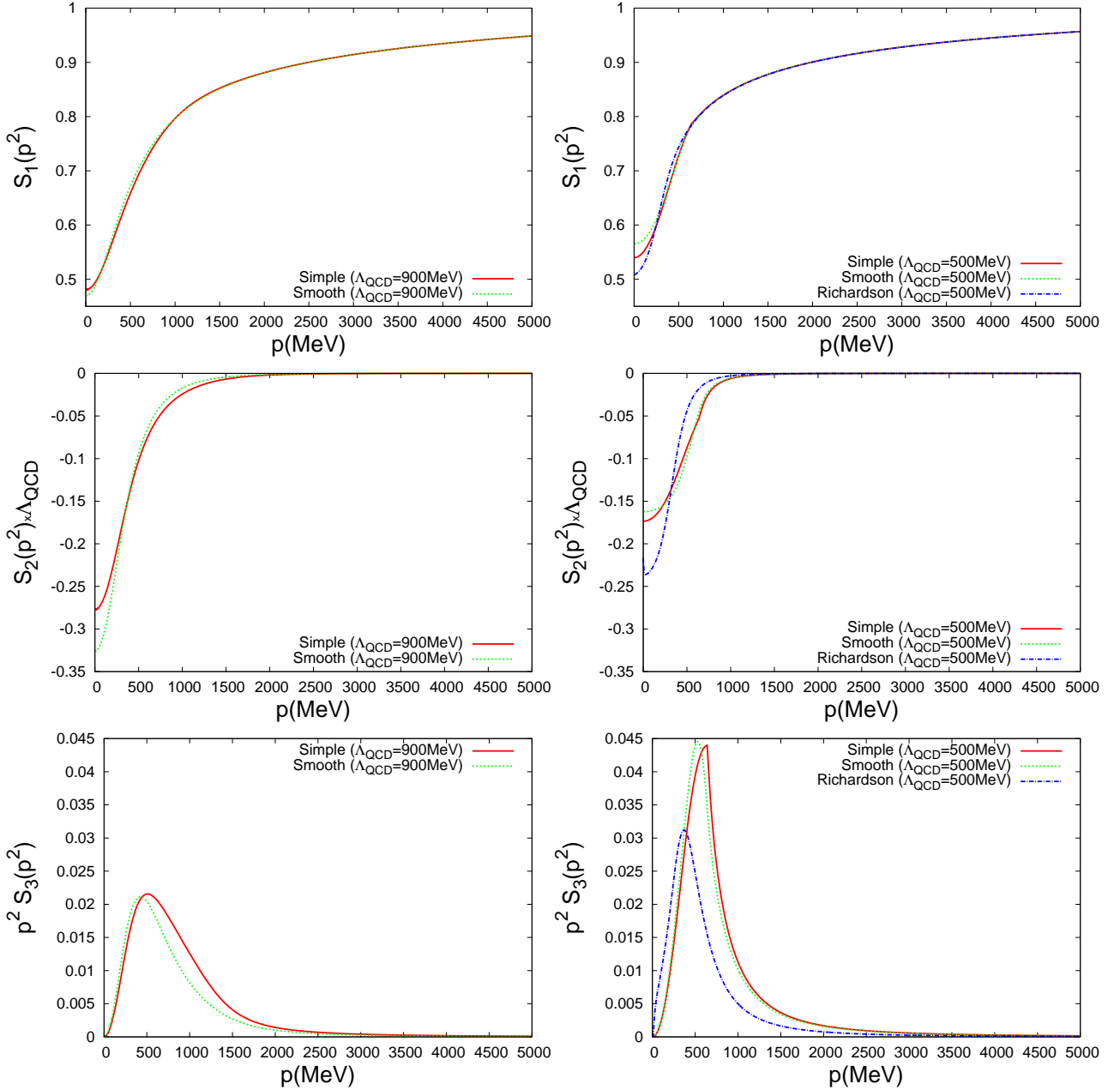}
\caption{\label{fig:tensor_charge_123}
The $S_1$, $S_2$, and $S_3$, functions (not renormalized) solved with the Schwinger-Dyson equation for the quark tensor charge with the integral cutoff $\Lambda = 20$ GeV.
The left column shows the results calculated without approximation, and the right column with the Higashijima-Miransky approximation.
The $S_3$ function was resized with $p^2$.
}
\end{figure*}

In Eq. (\ref{eq:tensorsde}), there are three relevant Lorentz structures: $\sigma^{\mu \nu}$, $\bigl\{ p\hspace{-.45em}/\, , \sigma^{\mu \nu} \bigr\} \, (\, \equiv p\hspace{-.45em}/\, \sigma^{\mu \nu} +\sigma^{\mu \nu} p\hspace{-.45em}/\, \, )$, and $\sigma^{\mu \rho} p_\rho p^\nu -\sigma^{\nu \rho} p_\rho p^\mu$.
The dynamical tensor charge is thus written as 
\begin{eqnarray}
\Sigma^{\mu \nu} (p) 
&\equiv &
S_1 (p^2) \sigma^{\mu \nu}
+S_2 (p^2) 
\bigl\{ p\hspace{-.45em}/\, , \sigma^{\mu \nu} \bigr\}
\nonumber\\
&&
+S_3 (p^2) (\sigma^{\mu \rho} p_\rho p^\nu -\sigma^{\nu \rho} p_\rho p^\mu )
\, .
\label{eq:tensorstructure}
\end{eqnarray}
The SD equation (\ref{eq:tensorsde}) can thus be rewritten in a set of integral equations with the $S_1 (p^2)$, $S_2 (p^2)$, and $S_3 (p^2)$ functions.
The zero momentum point of the $S_1$ function indicates the ratio between the tensor charges  of the dressed and bare quarks (it will be called simply ``quark tensor charge'' from now on).
After some algebra, we find the following set of integral equations:
\begin{widetext}
\begin{eqnarray}
S_1 (p_E^2)
&=&
1+
\frac{C_2 (N_c)}{3\pi^2} \int_0^\Lambda \hspace{-0.7em} k_E^3 dk_E
\int_0^\pi \hspace{-0.5em} \sin^2 \theta d\theta \frac{\alpha_s [(p_E - k_E)^2]}{\left[ k_E^2 +\Sigma^2 (k_E^2) \right]^2} Z^2 (k_E^2)
\nonumber\\
&& \hspace{5em}
\times \Biggl\{ 
S_1 (k_E^2) \Biggl[ \Biggl( \frac{\Sigma^2 (k_E^2)}{p_E^2} -1 \Biggr) \Biggl( 1+ \frac{(p_E^2 - k_E^2 )^2}{(p_E -k_E)^4} \Biggr) 
+\frac{\frac{\Sigma^2 (k_E^2)}{p_E^2} (p_E^2-2k_E^2 ) +2p_E^2-k_E^2 }{(p_E -k_E)^2}
\Biggr]
\nonumber\\
&& \hspace{7em}
+2 S_2 (k_E^2) \Sigma (k_E^2) \Biggl[ -\Biggl( 1+\frac{k_E^2}{p_E^2} \Biggr) \Biggl( 1+ \frac{(p_E^2 - k_E^2 )^2}{(p_E -k_E)^4} \Biggr) 
+2\frac{p_E^2 -k_E^2 +\frac{k_E^4}{p_E^2} }{(p_E -k_E)^2}
\Biggr]
\nonumber\\
&& \hspace{7em}
-\frac{1}{2} S_3 (k_E^2) \left[ k_E^2 +\Sigma^2 (k_E^2) \right] 
\Biggl[ \Biggl( \frac{k_E^2}{p_E^2} -1 \Biggr) \Biggl( 1+ \frac{(p_E^2 - k_E^2 )^2}{(p_E -k_E)^4} \Biggr) 
+2\frac{p_E^2 -\frac{k_E^4}{p_E^2} }{(p_E -k_E)^2} \Biggr]
\ \Biggr\}
,
\label{eq:sdes1}
\\
S_2(p_E^2)
&=&
\frac{C_2 (N_c)}{3\pi^2 p_E^2 }
\int_0^\Lambda \hspace{-0.7em} k_E^3 dk_E \int_0^\pi \hspace{-0.5em} \sin^2 \theta d\theta
\frac{\alpha_s [(p_E -k_E)^2]}{\left[ k_E^2 +\Sigma^2 (k_E^2) \right]^2} Z^2 (k_E^2)
\cdot
\left[
2 -\frac{5}{2} \frac{p_E^2 +k_E^2}{(p_E -k_E)^2} +\frac{1}{2} \frac{(p_E^2 -k_E^2)^2}{(p_E -k_E)^4}
\right]
\nonumber\\
&& \hspace{21.5em} \times
\Biggl\{
\Sigma (k_E^2) S_1 (k_E^2) 
-\left[ k_E^2 - \Sigma^2 (k_E^2) \right] S_2 (k_E^2)
\Biggr\}
,
\label{eq:sdes2}
\\
S_3(p_E^2)
&=&
\frac{C_2 (N_c)}{3\pi^2 p_E^2 }
\int_0^\Lambda \hspace{-0.7em} k_E^3 dk_E \int_0^\pi \hspace{-0.5em} \sin^2 \theta d\theta
\frac{\alpha_s [(p_E -k_E)^2]}{\left[ k_E^2 +\Sigma^2 (k_E^2) \right]^2} Z^2 (k_E^2)
\cdot
\left[
1 + \frac{p_E^2 -2k_E^2}{(p_E -k_E)^2} + \frac{(p_E^2 -k_E^2)^2}{(p_E -k_E)^4}
\right]
\nonumber\\
&& \hspace{10em} \times
\Biggl\{
2 \frac{\Sigma^2 (k_E^2)}{p_E^2} S_1 (k_E^2) 
-4\Sigma (k_E^2) \frac{k_E^2}{p_E^2} S_2 (k_E^2)
-\left[ k_E^2 + \Sigma^2 (k_E^2) \right] \frac{k_E^2}{p_E^2} S_3 (k_E^2)
\Biggr\}
,
\label{eq:sdes3}
\end{eqnarray}
For the derivation of the above integral equations, see the Appendix.
The result of the SD equation for the quark tensor charge is plotted in Fig. \ref{fig:tensor_charge_123}.

Applying the Higashijima-Miransky approximation (\ref{eq:Higashijima-Miransky}) to the quark tensor charge SD equations (\ref{eq:sdes1}), (\ref{eq:sdes2}), and (\ref{eq:sdes3}), we obtain
\begin{eqnarray}
S_1 (p_E^2)
&\approx &
1+
\frac{C_2(N_c)}{2\pi}
\int_{p_E}^\Lambda \hspace{-0.7em} k_E dk_E
\frac{\alpha_s [{\rm max}(p_E^2,k_E^2)]}{\left[ k_E^2 +\Sigma^2 (k_E^2) \right]^2} 
(p_E^2 - k_E^2)
\Biggl\{
S_1(k_E^2)
+2\Sigma (k_E^2) S_2(k_E^2)
-\frac{1}{2} \left[ k_E^2 + \Sigma^2 (k_E^2) \right] S_3 (k_E^2)
\Biggr\}
\nonumber\\
&& \hspace{1em}
+
\frac{C_2(N_c)}{2\pi}
\int_0^{p_E} \hspace{-1em} k_E dk_E
\frac{\alpha_s [{\rm max}(p_E^2,k_E^2)]}{\left[ k_E^2 +\Sigma^2 (k_E^2) \right]^2} 
(p_E^2 - k_E^2)
\frac{k_E^4}{p_E^4}
\nonumber\\
&& \hspace{10em} \times
\Biggl\{
\frac{\Sigma^2 (k_E^2)}{k_E^2} S_1(k_E^2)
-2\Sigma (k_E^2) S_2(k_E^2)
-\frac{1}{2} \left[ k_E^2 + \Sigma^2 (k_E^2) \right] S_3 (k_E^2)
\Biggr\}
,
\label{eq:sdehm1}
\end{eqnarray}
\begin{eqnarray}
S_2(p_E^2)
&\approx &
\frac{C_2(N_c)}{2\pi}
\int_{p_E}^\Lambda \hspace{-0.7em} k_E dk_E
\frac{\alpha_s [{\rm max}(p_E^2,k_E^2)]}{\left[ k_E^2 +\Sigma^2 (k_E^2) \right]^2} 
\Biggl\{
-\Sigma (k_E^2) S_1 (k_E^2) 
+\left[ k_E^2 - \Sigma^2 (k_E^2) \right] S_2 (k_E^2)
\Biggr\}
\nonumber\\
&&+
\frac{C_2(N_c)}{2\pi}
\int_0^{p_E} \hspace{-1em} k_E dk_E
\frac{\alpha_s [{\rm max}(p_E^2,k_E^2)]}{\left[ k_E^2 +\Sigma^2 (k_E^2) \right]^2} 
\cdot\frac{k_E^4}{p_E^4}
\Biggl\{
-\Sigma (k_E^2) S_1 (k_E^2) 
+\left[ k_E^2 - \Sigma^2 (k_E^2) \right] S_2 (k_E^2)
\Biggr\}
\, , \ \ 
\label{eq:sdehm2}
\\
S_3 (p_E^2)
&\approx &
\frac{C_2(N_c)}{2\pi}
\int_0^{p_E} \hspace{-1em} k_E dk_E
\frac{\alpha_s [{\rm max}(p_E^2,k_E^2)]}{\left[ k_E^2 +\Sigma^2 (k_E^2) \right]^2} 
(p_E^2 - k_E^2)
\frac{k_E^4}{p_E^6} 
\nonumber\\
&& \hspace{10em} \times
 \Biggl\{
2\frac{\Sigma^2 (k_E^2)}{k_E^2} S_1 (k_E^2) 
-4\Sigma (k_E^2 ) S_2 (k_E^2)
-\left[ k_E^2 +\Sigma^2 (k_E^2) \right] S_3 (k_E^2) 
\Biggr\}
\, .
\label{eq:sdehm3}
\end{eqnarray}
\end{widetext}
Here we note that the quark wave function renormalization factor was set to 1, since we have solved the SD equation of the quark propagator with the Higashijima-Miransky approximation (\ref{eq:Higashijima-Miransky}) to obtain the self-energy of the quark.
The result of the SD equation for the quark tensor charge in the Higashijima-Miransky approximation with three different running couplings is plotted in Fig. \ref{fig:tensor_charge_123}.

\section{\label{sec:analysis}Analysis and discussion}

The solution of the SD equation for the quark tensor charge (Fig. \ref{fig:tensor_charge_123}) shows a similar shape among different RG improved strong couplings (simple IR regularization, smooth IR regularization, and Richardson Ansatz), which suggests the good description of the quark tensor charge within this framework.
The quark tensor charge calculated in the Higashijima-Miransky approximation becomes larger than the result without it.
Since the results are similar in the simple, smooth, and Richardson cases, the confinement effect, which is phenomenologically introduced in the Richardson Ansatz, is expected to be small at least within the Higashijima-Miransky approximation.

We must note, however, that the $S_1$, $S_2$, and $S_3$ functions obtained after solving Eqs. (\ref{eq:sdes1}), (\ref{eq:sdes2}), and (\ref{eq:sdes3}) are dependent on the cutoff $\Lambda$, and we need to renormalize the tensor charge at some fixed scale.
To renormalize the tensor charge $S_1 (0)$ at some renormalization point $\mu$, we use the formula \cite{tensorrenormalization}
\begin{equation}
S_1 (0)_\mu 
=
\left( \frac{\alpha_s (\Lambda^2 )}{\alpha_s (\mu^2 ) } \right)^{-\frac{C_2 (N_c)}{16\pi^2 \beta_0}}
S_1 (0)_\Lambda
\, ,
\end{equation}
where $S_1 (0)_\mu$ is the renormalized tensor charge and $S_1 (0)_\Lambda$ is the tensor charge given as the solution of the cutoff ($\Lambda$) dependent SD equation.
The exponent is $-\frac{C_2 (N_c)}{16\pi^2 \beta_0}=-\frac{4}{27}$ for $N_c = 3$ and $N_f =3$.
This renormalization of $S_1 (0)$ obtained from the SD equation [Eqs. (\ref{eq:sdes1}), (\ref{eq:sdes2}), and (\ref{eq:sdes3})] shows a very good stability against the change of the cutoff $\Lambda$ (see Tables \ref{table:stability} and \ref{table:stability_hm}).
This formula is also consistent with the analysis of the running of the Wilson coefficient of the quark EDM \cite{degrassi} (note that, in that analysis, the operator involves the current quark mass, which shifts the exponent by $\frac{12}{27}$ for $N_f =3$).
From the above formula, we obtain the renormalized tensor charge at $\mu =2$ GeV:
\begin{eqnarray}
S_1 (0)_{\mu =2\, {\rm GeV}} &=& 0.588 \ \ \mbox{(Simple IR regularization)} \, ,
\nonumber\\
S_1 (0)_{\mu =2\, {\rm GeV}} &=& 0.575 \ \ \mbox{(Smooth IR regularization)} \, .
\nonumber\\
\end{eqnarray}
With the Higashijima-Miransky approximation, we obtain
\begin{eqnarray}
S_1 (0)_{\mu =2\, {\rm GeV}} &=& 0.624 \ \ \mbox{(Simple IR regularization)} \, ,
\nonumber\\
S_1 (0)_{\mu =2\, {\rm GeV}} &=& 0.653 \ \ \mbox{(Smooth IR regularization)} \, ,
\nonumber\\
S_1 (0)_{\mu =2\, {\rm GeV}} &=& 0.588 \ \ \mbox{(Richardson Ansatz)} \, .
\end{eqnarray}
We see that the renormalized $S_1$(0) is smaller than 1.
This fact shows that the tensor charge of the dressed quark is suppressed compared with the bare quark contribution by the gluon dressing of the vertex.

\begin{table}
\caption{The stability of the tensor charge in the change of the integral cutoff $\Lambda$.
The tensor charge was calculated with the simple IR regularization.
The renormalization point was fixed to $\mu = 2$ GeV.
The renormalization of the chiral condensate is also shown to emphasize the stability.
}
\begin{ruledtabular}
\begin{tabular}{ccccc}
$\Lambda$ & 4 GeV & 20 GeV & 100 GeV & 1 TeV \\
\hline
$S_1(0)_\Lambda$ & 0.542 & 0.480 & 0.450 & 0.424 \\
$S_1(0)_\mu$ & 0.594 & 0.588 & 0.586 & 0.584 \\
$\langle \bar q q \rangle_\Lambda$ & $-(296)^3$ & $-(306)^3$ & $-(316)^3$ & $-(328)^3$ \\
$\langle \bar q q \rangle_\mu$ & $-(270)^3$ & $-(250)^3$ & $-(243)^3$ & $-(238)^3$ \\
\end{tabular}
\end{ruledtabular}
\label{table:stability}
\end{table}

\begin{table}
\caption{The tensor charge in the change of the integral cutoff $\Lambda$ obtained with the Higashijima-Miransky approximation.
The setup is the same as for Table \ref{table:stability}.
}
\begin{ruledtabular}
\begin{tabular}{ccccc}
$\Lambda$ & 4 GeV & 20 GeV & 100 GeV & 1 TeV \\
\hline
$S_1(0)_\Lambda$ & 0.589 & 0.540 & 0.511 & 0.484 \\
$S_1(0)_\mu$ & 0.626 & 0.624 & 0.623 & 0.623 \\
$\langle \bar q q \rangle_\Lambda$ &$-(256)^3$ & $-(281)^3$ & $-(297)^3$ & $-(314)^3$ \\
$\langle \bar q q \rangle_\mu$ &$-(241)^3$ & $-(243)^3$ & $-(244)^3$ & $-(244)^3$ \\
\end{tabular}
\end{ruledtabular}
\label{table:stability_hm}
\end{table}

If we associate the dressed dynamical quark with the constituent quark, our result can be combined with the nonrelativistic constituent quark model prediction of the quark tensor charge in the nucleon
\begin{eqnarray}
\delta u &=& \frac{4}{3} S_1 (0)_\mu \simeq 0.8\, ,
\nonumber\\
\delta d &=& -\frac{1}{3} S_1 (0)_\mu \simeq -0.2 \, ,
\, .
\label{eq:(15)}
\end{eqnarray}
In the above derivation, it is, of course, assumed that the nucleon is composed of three constituent valence quarks with negligible spin dependent many-body interactions.
The suppression of the tensor charge agrees qualitatively with the results obtained from the extraction in the collinear factorization approach [see Eq. (\ref{eq:extraction})] and from the  lattice QCD calculations [see Eq. (\ref{eq:latticeqcd})].
Additional suppression of the tensor charge may occur due to the many-body effect, but this topic is beyond the scope of this paper.
It should be noted that the sea quark contribution is small, since the tensor charge of the antiquarks has opposite sign.
This fact is in contrast to the quark-spin distribution $g_1 (x)$ which receives contribution from both quarks and antiquarks with the same sign.
The smallness of the sea quark effect to the tensor charge is also consistent with the lattice QCD results.
We should also add that the dressed quark tensor charge has a small dependence on the scale parameter $\Lambda_{\rm QCD}$.
We show the coefficient $S_1 (0)$ for several values of $\Lambda_{\rm QCD}$ in Table \ref{table:lambdaqcddependence}.
This stability is due to the fact that the $S_1 (0)$ is a dimensionless number.
\begin{table}
\caption{The quark tensor charge obtained with several $\Lambda_{\rm QCD}$.
The renormalization point was fixed to $\mu = 2$ GeV.
}
\begin{ruledtabular}
\begin{tabular}{ccccc}
$\Lambda_{\rm QCD}$ & 200 MeV & 500 MeV & 900 MeV & 1 GeV \\
\hline
$S_1(0)_\mu$ & 0.500 & 0.541 & 0.588 & 0.600 \\
\end{tabular}
\end{ruledtabular}
\label{table:lambdaqcddependence}
\end{table}

Let us derive the contribution of the quark EDM to the nucleon EDM within the above simple model assumption.
By combining the simple constituent quark model with our result, we obtain
\begin{equation}
d_n \sim
0.8  d_d^{(\mu)}
- 0.2 d_u^{(\mu)} 
\, .
\label{eq:nedm}
\end{equation}
We must note that the quark EDM is not a renormalization group invariant quantity.
In this case, the bare quark EDMs $d_d^{(\mu)}$ and $d_u^{(\mu)}$ are defined at the renormaliztion point of our discussion, i.e. at $\mu = 2$ GeV.
To relate the prediction of the quark EDMs defined, for example, at $\mu_S = 1$ TeV, we need to connect them with the renormalization group running of the EDM operators \cite{degrassi}
\begin{equation}
d_q^{(\mu)} 
=
\left( \frac{\alpha_s (\mu_S^2 )}{\alpha_s (\mu^2 ) } \right)^{\frac{C_2 (N_c)}{16\pi^2 \beta_0}}
d_q^{(\mu_S)} 
\, ,
\end{equation}
where $\frac{C_2 (N_c)}{16\pi^2 \beta_0} = \frac{4}{27}$.
The running of the quark EDM from $\mu_S = 1$ TeV to 2 GeV brings thus a suppression factor of $\sim$ 0.8.
We thus have
\begin{equation}
d_n \sim
{ 0.6}  d_d^{(\mu_S= 1\, {\rm TeV})}
- { 0.1} d_u^{(\mu_S= 1\, {\rm TeV})} 
\, .
\label{eq:nedmtev}
\end{equation}
It should be noted that, in the above discussion, we have not considered the other $CP$-odd quark and gluon level operators.
In general, these $CP$-odd operators can mix with each other when the operators are rescaled from the TeV scale to the hadronic scale \cite{degrassi,hisano}.

\begin{figure}[htb]
\includegraphics[width=6.1cm,angle=-90]{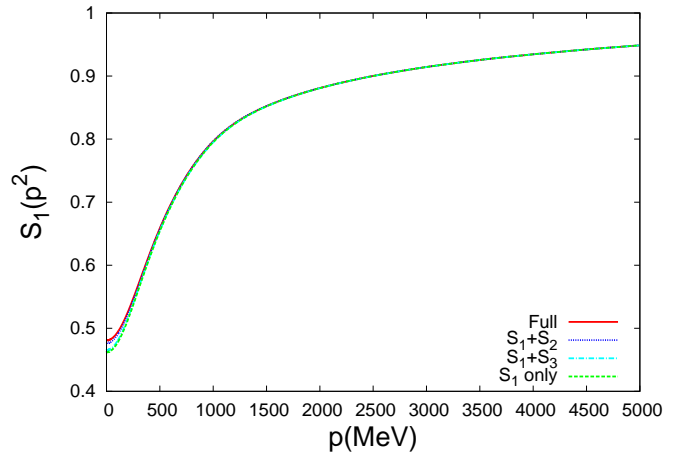}
\caption{\label{fig:tensor_s1_compare}
The $S_1$ function (not renormalized) obtained by solving the Schwinger-Dyson equation with $S_2$ and $S_3$ functions set to zero.
The $S_1$ function solved with the full contribution ($S_1$, $S_2$, and $S_3$) is also shown for comparison.
}
\end{figure}

In the formalism we have adopted, it is possible to change the input parameters and the self-energy function we have obtained in the intermediate steps, and this fact is an important advantage of the SD formalism.
We first tested the contribution of the $S_1$, $S_2$, and $S_3$ functions doing a fictitious manipulation by setting $S_2 (p^2) =0$ or/and $S_3 (p^2)=0$ in solving the SD equations (\ref{eq:sdes1}), (\ref{eq:sdes2}), and (\ref{eq:sdes3}).
The result is plotted in Fig. \ref{fig:tensor_s1_compare}.
We see that the solution of the SD equation with and without the contribution from $S_2$, and $S_3$ functions are close within 3\%.
The qualitative features are very similar.
It can be also seen that the effect from $S_2$ is more important than $S_3$.
This result suggests that the extra powers of momenta $p$ (appearing in $\bigl\{ \, p\hspace{-.45em}/\, , \sigma^{\mu \nu} \bigr\} $ and $\sigma^{\mu \rho} p_\rho p^\nu -\sigma^{\nu \rho} p_\rho p^\mu$) work as a suppression factor. 
This shows that the leading contribution to the SD equation of the quark tensor charge is given by the $S_1$ function and that the omission of $S_2$ and $S_3$ functions is a relatively good approximation.

\begin{figure}[htb]
\includegraphics[width=6.1cm,angle=-90]{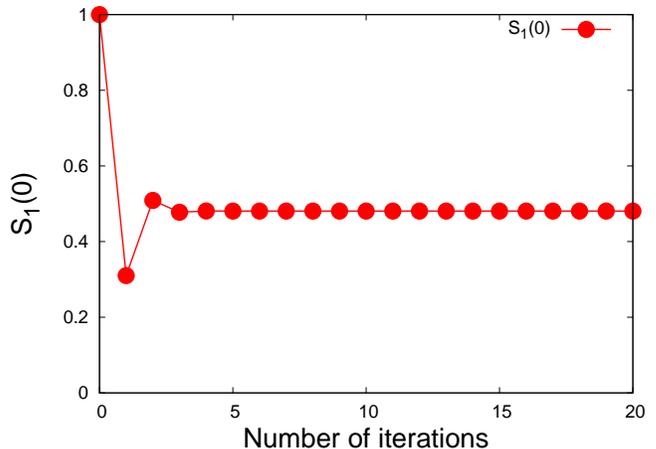}
\caption{\label{fig:tensor_s1(0)}
The convergence of $S_1$ function (not renormalized) at the origin in the number of iterations of the Schwinger-Dyson equation with the initial conditions $S_1(p^2) = 1$, $S_2(p^2) = 0$, and $S_3(p^2) = 0$.
}
\end{figure}

\begin{figure*}[htb]
\includegraphics[width=18cm]{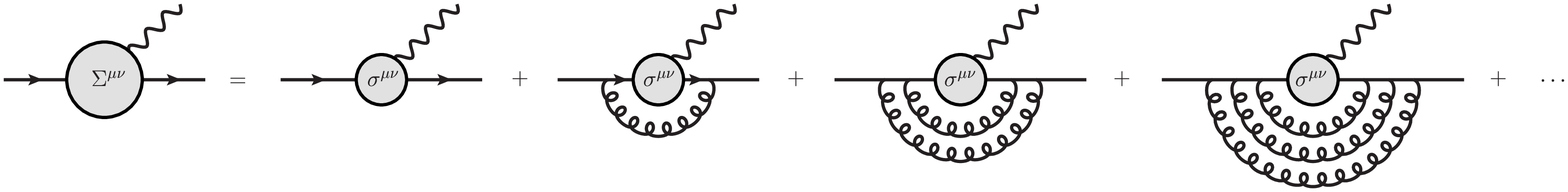}
\caption{\label{fig:tensor_SD_eq_expand}
Expansion of the Schwinger-Dyson equation for the quark tensor charge.
Each iteration gives the perturbative truncated contribution to the corresponding order.
}
\end{figure*}

\begin{figure}[htb]
\includegraphics[width=8.6cm]{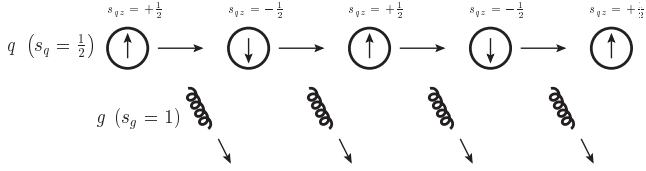}
\caption{\label{fig:quark_spin}
The schematic picture of the quark spin flip with the gluon emission.
}
\end{figure}

We now try to understand the suppression of the quark tensor charge with the gluon vertex dressing.
Let us first see the quark tensor charge obtained after few iterations.
The quark tensor charge $S_1 (0)$ calculated after each iteration is shown in Fig. \ref{fig:tensor_s1(0)}.
In our calculation of the SD equation, we have taken as the initial condition $S_1(p^2) = 1$, $S_2(p^2) = 0$, and $S_3(p^2) = 0$ and iteratively substituted the left-hand sides of Eqs. (\ref{eq:sdes1}), (\ref{eq:sdes2}), and (\ref{eq:sdes3}) to their right-hand sides.
This procedure can be seen as a sort of perturbative truncation, in which the number of the iteration corresponds to the order of perturbation (see Fig. \ref{fig:tensor_SD_eq_expand}).
The initial value $S_1(p^2) = 1$ is the bare quark tensor charge.
From Fig. \ref{fig:tensor_s1(0)}, we can see that the tensor charge converges by oscillating around the true tensor charge.
This means that the gluon dressed tensor vertex is decomposed into terms that change their sign alternatively in the perturbative expansion.
This fact can be understood as follows.
The tensor charge is given by the spin of the quark in the nonrelativistic limit, so the gluon emission of the quark changes the sign of the tensor charge, since the angular momenta of the quark and the gluon are, respectively, $s_q=\frac{1}{2}$ and $s_g=1$.
The above description is illustrated schematically in Fig. \ref{fig:quark_spin}.
As the external field can only probe the tensor charge (spin) of the quark, the superposition of the contribution of each order is always smaller than the bare contribution.

\begin{figure}[htb]
\includegraphics[width=6.1cm,angle=-90]{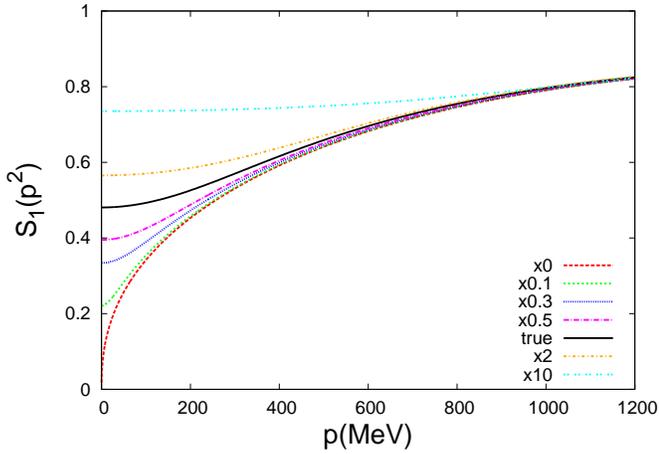}
\caption{\label{fig:tensor_resize}
The $S_1$ function (not renormalized) calculated with resized self-energy.
}
\end{figure}

The suppression of the quark tensor charge by the quark spin flip can be confirmed by artificially manipulating the self-energy of the quark.
The self-energy of the quark can be seen as its mass, so the spin flip of the quark should be suppressed when the quark becomes heavier.
The $S_1$ function calculated with the resized quark self-energy is plotted in Fig. \ref{fig:tensor_resize}.
We can see that the quark tensor charge approaches 1 when the self-energy is taken larger.
This result is consistent with our description: as the quark spin flip is suppressed for the heavy dressed quark, the contribution from the higher-order dressed tensor vertex becomes smaller, and the dressed quark tensor charge keeps a value close to the bare quark one.
On the contrary, the quark tensor charge vanishes when the quark becomes lighter (with smaller self-energy), since the spin flip becomes important so that the tensor charge is averaged at zero.

\section{\label{sec:summary}Summary}

In this paper, we have calculated the tensor charge of the quark in the QCD-like theory with the Landau gauge using the SD formalism with three different running couplings.
As a result, the quark tensor charge is suppressed by a factor of { $\sim 0.6$} compared to the bare quark contribution.
By combining with the nonrelativistic constituent quark model, the quark tensor charge in nucleon is given as { $\delta u \simeq 0.8$} and { $\delta d \simeq -0.2 $} when the renormalization scale is taken as $\mu = 2$ GeV.

Our result agrees qualitatively with the results obtained from the extraction of the tensor charge within the collinear factorization approach based on the experimental data and also with those given by the first principle lattice QCD studies, both suggesting the suppression of the quark tensor charge in the nucleon.

The stability of the renormalized quark tensor charge in the change of the integral cutoff, which is a requirement of this framework, is also fulfilled for the calculations with and without the Higashijima-Miransky approximation.
We have also shown that the phenomenological strong coupling of Richardson Ansatz can be used with the Higashijima-Miransky approximation, since the latter works as a regularization against the singularity $p-k=0$.

The result of our study gives also the contribution of the quark EDM to the neutron EDM.
The neutron EDM receives a contribution from the quark EDM defined at $\mu_S = 1$ TeV as { $d_n \sim 0.6  d_d^{(\mu_S= 1\, {\rm TeV})} - 0.1 d_u^{(\mu_S= 1\, {\rm TeV})} $.}

Through the analysis, we concluded two important results.
First, the dominant contribution of the dressed tensor charge is given by the $S_1$ function, the coefficient of the $\sigma^{\mu \nu}$ Dirac matrix.
Second, we have deduced that the suppression of the quark tensor charge is due to the superposition of the spin flipped states occurring in the gluon emission.
The gluon dressing of the vertex thus plays a crucial role in the suppression of the quark tensor charge, and this partially explains the deviation of the results suggested by the collinear factorization approach and lattice QCD from that given in the nonrelativistic constituent quark model.

We must however note that we have only discussed the single quark contribution to the nucleon tensor charge.
The remaining effect to the nucleon tensor charge should be investigated in the viewpoint of the many-body physics of partons.
It is actually suggested that the orbital angular momentum of the nucleonic partons carries a large fraction of the nucleon spin \cite{emc,angularmomentum}, and it is strongly probable that the bound state effect of the quark in nucleon contributes to the { modification} of the quark tensor charge.
The study of the many-body effect will be the subject of the next work.
Here, we briefly give the prospect for the improvement.
The first possibility is to include the quark in the nucleon with the quark model.
The second possibility is to the include the dressed tensor vertex in the SD equation of the quark-diquark bound state, which was investigated in Refs. \cite{hecht,emformfactor}.
The ideal way of the SD formalism is to formulate and calculate the relativistic Faddeev equation for the three-quark state \cite{faddeev}.

\begin{acknowledgments}

NY thanks T. Hatsuda, Y. Hatta, and H. Iida for useful discussion and comments.
He thanks also T. Ichihara for technical help.
This work is in part supported by the Grant for Scientific Research [Priority Areas ``New Hadrons'' (E01:21105006), (C) No.23540306] from the Ministry of Education, Culture,Science and Technology (MEXT) of Japan.

\end{acknowledgments}

\appendix

\onecolumngrid

\section{\label{sec:tensor_derivation}Detailed calculation of the Schwinger-Dyson equation for the quark tensor charge}

The Schwinger-Dyson equation for the quark tensor charge [Eq. (\ref{eq:tensorsde})] is rewritten as
\begin{eqnarray}
\Sigma^{\mu \nu} (p)
&=&
\sigma^{\mu \nu}+
i\int \frac{d^4k}{4\pi^3}
\cdot \frac{\alpha_s [(p-k)^2] Z^2(k^2)}{\left[ k^2 -\Sigma^2 (k^2) \right]^2} 
\cdot \frac{-1}{(p-k)^2} \left[ g^{\rho \lambda} - \frac{(p-k)^\rho (p-k)^\lambda}{(p-k)^2} \right]
C_2 (N_c)
\nonumber\\
&& \hspace{5em} \times
\gamma_\rho
\left[ k\hspace{-.45em}/\, + \Sigma (k^2)\right]
\left[ 
S_1(k^2) \sigma^{\mu \nu} 
+S_2 (k^2) 
\bigl\{ k\hspace{-.45em}/\, , \sigma^{\mu \nu} \bigr\}
+S_3 (k^2) (\sigma^{\mu \rho} k_\rho k^\nu -\sigma^{\nu \rho} k_\rho k^\mu )
\right]
\left[ k\hspace{-.45em}/\, + \Sigma (k^2)\right] \gamma_\lambda
\, .
\nonumber\\
\label{eq:tensorsde2}
\end{eqnarray}

The Lorentz and Dirac structures of the term with $S_1 (k^2)$ of Eq. (\ref{eq:tensorsde2}) can be transformed as
\begin{eqnarray}
&&
\left[ g^{\rho \lambda} - \frac{(p-k)^\rho (p-k)^\lambda}{(p-k)^2} \right]
\gamma_\rho
\left[ k\hspace{-.45em}/\, + \Sigma \right]
 \sigma^{\mu \nu} 
\left[ k\hspace{-.45em}/\, + \Sigma \right] 
\gamma_\lambda
\nonumber\\
&=&
-(k^2 +\Sigma^2 ) \sigma^{\mu \nu} 
+ \Sigma \Bigl\{ \, 2k\hspace{-.45em}/\, +p\hspace{-.45em}/\, , \sigma^{\mu \nu} \Bigr\}
+2\left[ \sigma^{\mu \eta } p_\eta k^\nu -\sigma^{\nu \eta } p_\eta k^\mu  \right]
\nonumber\\
&&
- \frac{\Sigma}{(p-k)^2}
\left[
(k^2 - p^2) \Bigl\{ \, k\hspace{-.45em}/\, - p\hspace{-.45em}/\, , \sigma^{\mu \nu} \Bigr\}
-i(p^\mu -k^\mu ) \Bigl\{ \, \gamma^\nu , [p\hspace{-.45em}/\, , k\hspace{-.45em}/\, ] \, \Bigr\}
+i(p^\nu -k^\nu ) \Bigl\{ \, \gamma^\mu , [p\hspace{-.45em}/\, , k\hspace{-.45em}/\, ] \, \Bigr\}
\right]
\nonumber\\
&&
+\frac{2}{(p-k)^2} \Biggl\{ 
(p^2 +\Sigma^2 ) \left[ \sigma^{\mu \eta } (p-k)_\eta (p-k)^\nu -\sigma^{\nu \eta } (p-k)_\eta (p-k)^\mu \right]
\nonumber\\
&& \hspace{6em}
-(p^2 -k^2 )
\left[ \sigma^{\mu \eta } (p-k)_\eta p^\nu -\sigma^{\nu \eta } (p-k)_\eta p^\mu \right]
+ 2 \sigma^{\rho \eta } k_\rho p_\eta (k^\nu p^\mu -k^\mu p^\nu )
\Biggl\}
\, .
\label{eq:s1lorentz}
\end{eqnarray}
For simplicity, we have omitted the argument of the self-energy $\Sigma$.
Similarly, the Lorentz and Dirac structures of the term with $S_2 (k^2)$ can be obtained as
\begin{eqnarray}
&&
\left[ g^{\rho \lambda} - \frac{(p-k)^\rho (p-k)^\lambda}{(p-k)^2} \right]
\gamma_\rho
\left[ k\hspace{-.45em}/\, + \Sigma \right]
\left( k\hspace{-.45em}/\, \sigma^{\mu \nu} +\sigma^{\mu \nu} k\hspace{-.45em}/\, \right)
\left[ k\hspace{-.45em}/\, + \Sigma \right] 
\gamma_\lambda
\nonumber\\
&=&
-4 \Sigma k^2 \sigma^{\mu \nu}
+( k^2 + \Sigma^2 )\Bigl\{ \, 2k\hspace{-.45em}/\, +p\hspace{-.45em}/\, , \sigma^{\mu \nu} \Bigr\}
+4\Sigma ( \sigma^{\mu \rho}p_\rho k^\nu - \sigma^{\nu \rho } p_\rho k^\mu )
\nonumber\\
&&
- \frac{( k^2 + \Sigma^2 ) }{(p-k)^2} 
\left[
( p^2 -k^2) \Bigl\{ \, p\hspace{-.45em}/\, -k\hspace{-.45em}/\, , \sigma^{\mu \nu} \Bigr\}
- i(p-k)^\mu \Bigl\{ \, \gamma^\nu , [p\hspace{-.45em}/\, , k\hspace{-.45em}/\, ] \Bigr\}
+ i(p-k)^\nu \Bigl\{ \, \gamma^\mu , [p\hspace{-.45em}/\, , k\hspace{-.45em}/\, ] \Bigr\}
\right]
\nonumber\\
&&
+\frac{4\Sigma}{(p-k)^2} \Biggl[ \,
2k^2 \Bigl[ \sigma^{\mu \rho}(p-k)_\rho p^\nu -\sigma^{\nu \rho}(p-k)_\rho p^\mu \Bigr]
\nonumber\\
&& \hspace{6em}
-(p^2 + k^2) \Bigl[ \sigma^{\mu \rho}(p-k)_\rho k^\nu -\sigma^{\nu \rho}(p-k)_\rho k^\mu \Bigr]
+2\sigma^{\eta \rho}k_\eta p_\rho (p^\mu k^\nu - p^\nu k^\mu)
\Biggr]
\, ,
\label{eq:s2lorentz}
\end{eqnarray}
and the $S_3 (k^2)$ contribution as
\begin{eqnarray}
&&
\left[ g^{\rho \lambda} - \frac{(p-k)^\rho (p-k)^\lambda}{(p-k)^2} \right]
\gamma_\rho
\left[ k\hspace{-.45em}/\, + \Sigma \right]
\left( \sigma^{\mu \rho} k_\rho k^\nu - \sigma^{\nu \rho} k_\rho k^\mu \right)
\left[ k\hspace{-.45em}/\, + \Sigma \right] 
\gamma_\lambda
\nonumber\\
&=&
(k^2 -\Sigma^2 ) \left( \sigma^{\mu \rho} p_\rho k^\nu - \sigma^{\nu \rho} p_\rho k^\mu \right)
\nonumber\\
&&
-\frac{k^2 -\Sigma^2}{(p-k)^2}
\Biggl\{
(p^2 -k^2) \left[ \sigma^{\mu \rho} (p-k)_\rho k^\nu - \sigma^{\nu \rho} (p-k)_\rho k^\mu 
\right]
-2\sigma^{\rho \eta} k_\rho p_\eta (p^\mu k^\nu - p^\nu k^\mu)
\Biggr\}
\, .
\label{eq:s3lorentz}
\end{eqnarray}

In the transformation of the above equation, we have used the following identities:
\begin{eqnarray}
\gamma_\rho \sigma^{\mu \nu} \gamma^\rho
&=&
0 \, ,
\\
q\hspace{-.45em}/\, \sigma^{\mu \nu} q\hspace{-.45em}/\,
&=&
q^2 \sigma^{\mu \nu} -2 \sigma^{\mu \rho} q_\rho q^\nu +2 \sigma^{\nu \rho} q_\rho q^\mu \, ,
\label{eq:qsigmaq}
\\
\gamma_\rho \left( k\hspace{-.45em}/\, \sigma^{\mu \nu} +\sigma^{\mu \nu} k\hspace{-.45em}/\, \right) \gamma^\rho 
&=&
2 \left( k\hspace{-.45em}/\, \sigma^{\mu \nu} +\sigma^{\mu \nu} k\hspace{-.45em}/\, \right)
\, ,
\\
k\hspace{-.45em}/\, \left( k\hspace{-.45em}/\, \sigma^{\mu \nu} +\sigma^{\mu \nu} k\hspace{-.45em}/\, \right) k\hspace{-.45em}/\,
&=&
k^2 \left( k\hspace{-.45em}/\, \sigma^{\mu \nu} +\sigma^{\mu \nu} k\hspace{-.45em}/\, \right)
\, ,
\\
p\hspace{-.45em}/\, \left( k\hspace{-.45em}/\, \sigma^{\mu \nu} +\sigma^{\mu \nu} k\hspace{-.45em}/\, \right) p\hspace{-.45em}/\,
&=&
-p^2 \left( k\hspace{-.45em}/\, \sigma^{\mu \nu} +\sigma^{\mu \nu} k\hspace{-.45em}/\, \right)
+2(p\cdot k) \left( p\hspace{-.45em}/\, \sigma^{\mu \nu} +\sigma^{\mu \nu} p\hspace{-.45em}/\, \right)
\nonumber\\
&&
-ip^\mu \Bigl\{ \,  \gamma^\nu , [  p\hspace{-.45em}/\, , k\hspace{-.45em}/\, ]\, \, \Bigr\}
+ip^\nu \Bigl\{ \,  \gamma^\mu , [  p\hspace{-.45em}/\, , k\hspace{-.45em}/\, ]\, \, \Bigr\}
\label{eq:p(ksigma-sigmak)p}
\, .
\end{eqnarray}
We have also used the cyclic property $\Bigl\{ \,  \gamma^\mu , [ \gamma^\nu , \gamma^\rho ]\, \, \Bigr\} = \Bigl\{ \,  \gamma^\rho , [ \gamma^\mu , \gamma^\nu ]\, \, \Bigr\}$, which implies $\Bigl\{ \, p\hspace{-.45em}/\, , [  p\hspace{-.45em}/\, , k\hspace{-.4em}/\, ]\, \, \Bigr\} = \Bigl\{ \, k\hspace{-.4em}/\, , [  p\hspace{-.45em}/\, , p\hspace{-.45em}/\, ]\, \, \Bigr\} = 0$.

By substituting Eqs. (\ref{eq:s1lorentz}), (\ref{eq:s2lorentz}), and (\ref{eq:s3lorentz}) into Eq. (\ref{eq:tensorsde2}), we can further transform the integral equation (\ref{eq:tensorsde2}) as
\begin{eqnarray}
\Sigma^{\mu \nu} (p)
&=&
\sigma^{\mu \nu}
\nonumber\\
&&
-i \int \frac{d^4k}{4\pi^3}
\cdot \frac{\alpha_s [(p-k)^2] Z^2 (k^2)}{\left[ k^2 -\Sigma^2 (k^2) \right]^2} 
\cdot \frac{C_2 (N_c)}{(p-k)^2} \cdot S_1(k^2)
\nonumber\\
&& \hspace{5em} \times 
\Biggl\{
\frac{1}{3} \sigma^{\mu \nu} \left[ 
2p^2 -k^2 +\Sigma^2 (k^2) \left( 2\frac{k^2}{p^2} -1 \right) -\frac{p^2+ \Sigma^2 (k^2) }{p^2} \frac{(k^2-p^2)^2}{(p-k)^2} -\frac{p^2+\Sigma^2 (k^2)}{p^2} (p-k)^2
\right]
\nonumber\\
&& \hspace{6em}
+
\Sigma (k^2) \Bigl\{ \,p\hspace{-.45em}/\, , \sigma^{\mu \nu} \Bigr\}
\cdot \frac{1}{6 p^2}
\left[
5 (k^2+p^2) - \frac{(k^2-p^2)^2}{(k-p)^2} -4(p-k)^2
\right]
\nonumber\\
&& \hspace{6em}
+\frac{2}{3p^4} \Sigma^2 (k^2) (\sigma^{\mu \rho} p_\rho p^\nu -\sigma^{\nu \rho} p_\rho p^\mu )
\left[ 
(p^2 - 2k^2) + (p-k)^2 +\frac{(k^2 -p^2)^2}{(p-k)^2}
\right]
\ \Biggr\}
\nonumber\\
&&
-i \int \frac{d^4k}{4\pi^3}
\cdot \frac{\alpha_s [(p-k)^2] Z^2 (k^2)}{\left[ k^2 -\Sigma^2 (k^2) \right]^2} 
\cdot \frac{C_2 (N_c)}{(p-k)^2} \cdot S_2(k^2)
\nonumber\\
&& \hspace{5em} \times 
\Biggl\{
\frac{2}{3p^2} \Sigma (k^2) \sigma^{\mu \nu}
\left[ 
-(p^2 +k^2) \frac{(p^2-k^2)^2}{(p-k)^2} -(k^2 +p^2) (p-k)^2
+2(p^2+k^2)^2 -6k^2 p^2
\right]
\nonumber\\
&& \hspace{6.2em}
+
\left[ k^2 + \Sigma^2 (k^2) \right] \Bigl\{ \,p\hspace{-.45em}/\, , \sigma^{\mu \nu} \Bigr\}
\cdot \frac{1}{6 p^2}
\left[
5 (k^2+p^2) - \frac{(k^2-p^2)^2}{(k-p)^2} -4(p-k)^2
\right]
\nonumber\\
&& \hspace{6.2em}
+\frac{4k^2}{3p^4}\Sigma (k^2) (\sigma^{\mu \rho} p_\rho p^\nu -\sigma^{\nu \rho} p_\rho p^\mu )
\left[
\frac{(k^2- p^2)^2}{(p-k)^2}
+p^2 -2k^2
+(p-k)^2
\right]
\ \Biggr\}
\nonumber\\
&&
-i \int \frac{d^4k}{4\pi^3}
\cdot \frac{\alpha_s [(p-k)^2] Z^2 (k^2)}{ k^2 -\Sigma^2 (k^2)} 
\cdot \frac{C_2 (N_c)}{(p-k)^2} \cdot S_3(k^2)
\nonumber\\
&& \hspace{5em} \times 
\Biggl\{
\frac{p^2-k^2}{6p^2} \sigma^{\mu \nu}
\left[
-\frac{(k^2-p^2)^2}{(p-k)^2} +2(k^2+p^2) -(p-k)^2
\right]
\nonumber\\
&&\hspace{6.2em}
+\frac{k^2}{3p^4} (\sigma^{\mu \rho} p_\rho p^\nu - \sigma^{\nu \rho} p_\rho p^\mu )
\left[
2k^2 -p^2 -\frac{(p^2-k^2)^2}{(p-k)^2} -(p-k)^2
\right]
\ \Biggr\}
\, .
\label{eq:tensorsde3}
\end{eqnarray}
Here, we have used the formulae of the loop integral developed by Passarino and Veltman \cite{passarino-veltman} to reduce into a Lorentz scalar loop integral.
The rank-1 ($k^\mu$) integral can be reduced as
\begin{equation}
\int d^4 k \, F_1 (p,k) k^\mu
=
T_1 (p^2) p^\mu
\, ,
\end{equation}
where
\begin{equation}
T_1 (p^2) 
=
\int d^4 k \, F_1 (p, k ) \left[ \frac{p^2+k^2- (p-k)^2}{2p^2} \right]
=
\frac{1}{2}
\int d^4 k \, F_1 (p, k ) \left[ 1 +\frac{k^2}{p^2} - \frac{(p-k)^2}{p^2} \right]
\, .
\label{eq:t1}
\end{equation}
The rank-2 ($k^\mu k^\nu$) integral can be reduced as
\begin{equation}
\int d^4 k \, F_2 (p,k) k^\mu k^\nu
=
T_{00} (p^2) g^{\mu \nu}
+T_{11} (p^2) p^\mu p^\nu
\, ,
\end{equation}
where
\begin{eqnarray}
T_{00} (p^2) 
&=&
\frac{1}{3}
\int d^4 k \, F_2 (p,k) \, \left[ k^2 - \frac{1}{4} \frac{\left[ k^2 +p^2 -(p-k)^2 \right]^2}{p^2} \right]
\nonumber\\
&=&
\frac{1}{3}
\int d^4 k \, F_2 (p,k) \, \left[ \frac{1}{2} k^2 - \frac{1}{4} p^2 -\frac{1}{4} \frac{k^4}{p^2} +\frac{1}{2} \frac{(p^2+k^2)(p-k)^2}{p^2} -\frac{1}{4} \frac{(p-k)^4}{p^2} \right]
\, ,
\label{eq:t00}
\\
T_{11} (p^2) 
&=&
\frac{1}{3p^2}
\int d^4 k \, F_2 (p,k) \, \left[ \frac{\left[ k^2 +p^2 -(p-k)^2 \right]^2}{p^2} -k^2 \right]
\nonumber\\
&=&
\frac{1}{3}
\int d^4 k \, F_2 (p,k) \, \left[ 1+ \frac{k^2}{p^2} + \frac{k^4}{p^4} -2 \frac{(k^2+p^2)(p-k)^2}{p^4} +\frac{(p-k)^4}{p^4} \right]
\, .
\label{eq:t11}
\end{eqnarray}

By taking the trace after multiplying by $\sigma^{\mu \nu}$, Eq. (\ref{eq:tensorsde3}) can be rewritten as
\begin{eqnarray}
2 S_1 (p_E^2) - S_3 (p_E^2)p_E^2
&=&
2+
\frac{C_2 (N_c)}{12\pi^3}
\int d^4k_E
\cdot \frac{\alpha_s [(p_E -k_E)^2] Z^2 (k_E^2)}{\left[ k_E^2 +\Sigma^2 (k_E^2) \right]^2} 
\cdot
\left[
-1+ \frac{2 p_E^2 -k_E^2}{(p_E - k_E)^2}  - \frac{(p_E^2 -k_E^2)^2}{(p_E -k_E)^4}
\right]
\nonumber\\
&& \hspace{10em} \times
\Biggl[
2S_1(k_E^2)
+4\Sigma (k_E^2)
S_2(k_E^2)
-\left[ k_E^2 + \Sigma^2 (k_E^2) \right] 
S_3 (k_E^2)
\Biggr]
,
\label{eq:sdes1'}
\end{eqnarray}
where we have used the trace formulae
\begin{eqnarray}
{\rm Tr} \left[ \sigma^{\mu \nu} \sigma_{\mu \nu} \right] &=& 48 \ ,
\\
{\rm Tr} \left[ \sigma^{\mu \nu} ( \sigma_{\mu \rho} p^\rho p_\nu -\sigma_{\nu \rho} p^\rho p_\mu ) \right] &=& 24 p^2\ .
\label{eq:trace2}
\end{eqnarray}

By taking the trace after multiplying by $\left\{ p\hspace{-.45em}/\, , \sigma^{\mu \nu} \right\}$, Eq. (\ref{eq:tensorsde3}) can be rewritten as
\begin{eqnarray}
S_2(p_E^2)
&=&
\frac{C_2 (N_c)}{p_E^2 12\pi^3 }
\int d^4k_E
\cdot \frac{\alpha_s [(p_E -k_E)^2] Z^2 (k_E^2)}{\left[ k_E^2 +\Sigma^2 (k_E^2) \right]^2} 
\cdot
\left[
2 - \frac{5}{2} \frac{p_E^2 +k_E^2}{(p_E -k_E)^2} +\frac{1}{2} \frac{(p_E^2 -k_E^2)^2}{(p_E -k_E)^4}
\right]
\nonumber\\
&& \hspace{15em} \times
\Biggl[
\Sigma (k_E^2) S_1 (k_E^2) 
-\left[ k_E^2 - \Sigma^2 (k_E^2) \right] S_2 (k_E^2)
\Biggr]
\, .
\label{eq:sdes2'}
\end{eqnarray}
In deriving the above equation, we have used 
\begin{eqnarray}
{\rm Tr} \left[ \left\{ p\hspace{-.45em}/\, , \sigma^{\mu \nu} \right\} \sigma_{\mu \nu} \right] &=& 0 \ ,
\\
{\rm Tr} \left[ \left\{ p\hspace{-.45em}/\, , \sigma^{\mu \nu} \right\} \left\{ p\hspace{-.45em}/\, , \sigma_{\mu \nu} \right\}  \right] &=& 96p^2 \ ,
\\
{\rm Tr} \left[ \left\{ p\hspace{-.45em}/\, , \sigma^{\mu \nu} \right\} ( \sigma_{\mu \rho} p^\rho p_\nu -\sigma_{\nu \rho} p^\rho p_\mu ) \right] &=& 0\ .
\end{eqnarray}

By taking the trace after multiplying by $\sigma^{\mu \rho} p_\rho p^\nu -\sigma^{\nu \rho} p_\rho p^\mu $, Eq. (\ref{eq:tensorsde2}) can be rewritten as
\begin{eqnarray}
S_1 (p_E^2) - S_3 (p_E^2)p_E^2
&=&
1-
\frac{C_2(N_c)}{12\pi^3 p_E^2}
\int d^4k_E
\cdot \frac{\alpha_s [(p_E - k_E)^2] Z^2 (k_E^2)}{\left[ k_E^2 +\Sigma^2 (k_E^2) \right]^2} 
\nonumber\\
&& \hspace{2em} \times
\Biggl\{
S_1 (k_E^2)
\Biggl[
p_E^2 +\Sigma^2 ( k_E^2 ) 
+\frac{ p_E^2 (\Sigma^2 ( k_E^2 )-2p_E^2 ) +k_E^2 (p_E^2 -2 \Sigma^2 ( k_E^2 ) ) }{(p_E -k_E)^2}
\nonumber\\
&& \hspace{20em}
+ (p_E^2 +\Sigma^2 ( k_E^2 ) ) \frac{(p_E^2 -k_E^2)^2}{(p_E -k_E)^4}
\Biggr]
\nonumber\\
&& \hspace{4em} 
+ 2 \Sigma (k_E^2 ) ( p_E^2 -k_E^2) S_2 (k_E^2) 
\left[
1-2\frac{p_E^2 +k_E^2}{(p_E -k_E)^2} +\frac{(p_E^2 -k_E^2)^2}{(p_E -k_E)^4}
\right]
\nonumber\\
&& \hspace{4em} 
-\frac{1}{2} \left[ k_E^2 +\Sigma^2 (k_E^2) \right]  S_3 (k_E^2) 
\Biggl[
p_E^2 + k_E^2 -2 \frac{p_E^4- p_E^2 k_E^2 + k_E^4}{(p_E -k_E)^2} 
\nonumber\\
&& \hspace{22em}
+ (p_E^2 + k_E^2) \frac{(p_E^2 -k_E^2)^2}{(p_E -k_E)^4}
\Biggr] \ 
\Biggr\}
\, .
\label{eq:sdes3'}
\end{eqnarray}
In the derivation of the above equation, we have used Eq. (\ref{eq:trace2}) and
\begin{equation}
{\rm Tr} \left[ ( \sigma^{\mu \rho} p_\rho p^\nu -\sigma^{\nu \rho} p_\rho p^\mu ) ( \sigma_{\mu \eta} p^\eta p_\nu -\sigma_{\nu \eta} p^\eta p_\mu ) \right] = 24 p^4\ .
\end{equation}

By equating Eqs. (\ref{eq:sdes1'}) and (\ref{eq:sdes3'}), we obtain the system of integral equations for $S_1$, $S_2$, and $S_3$
\begin{eqnarray}
S_1 (p_E^2)
&=&
1+
\frac{C_2 (N_c)}{3\pi^2} \int_0^\Lambda \hspace{-0.7em} k_E^3 dk_E
\int_0^\pi \hspace{-0.5em} \sin^2 \theta d\theta \frac{\alpha_s [(p_E - k_E)^2]}{\left[ k_E^2 +\Sigma^2 (k_E^2) \right]^2} Z^2 (k_E^2)
\nonumber\\
&& \hspace{5em}
\times \Biggl\{ 
S_1 (k_E^2) \Biggl[ \Biggl( \frac{\Sigma^2 (k_E^2)}{p_E^2} -1 \Biggr) \Biggl( 1+ \frac{(p_E^2 - k_E^2 )^2}{(p_E -k_E)^4} \Biggr) 
+\frac{\frac{\Sigma^2 (k_E^2)}{p_E^2} (p_E^2-2k_E^2 ) +2p_E^2-k_E^2 }{(p_E -k_E)^2}
\Biggr]
\nonumber\\
&& \hspace{7em}
+2 S_2 (k_E^2) \Sigma (k_E^2) \Biggl[ -\Biggl( 1+\frac{k_E^2}{p_E^2} \Biggr) \Biggl( 1+ \frac{(p_E^2 - k_E^2 )^2}{(p_E -k_E)^4} \Biggr) 
+2\frac{p_E^2 -k_E^2 +\frac{k_E^4}{p_E^2} }{(p_E -k_E)^2}
\Biggr]
\nonumber\\
&& \hspace{7em}
-\frac{1}{2} S_3 (k_E^2) \left[ k_E^2 +\Sigma^2 (k_E^2) \right] 
\Biggl[ \Biggl( \frac{k_E^2}{p_E^2} -1 \Biggr) \Biggl( 1+ \frac{(p_E^2 - k_E^2 )^2}{(p_E -k_E)^4} \Biggr) 
+2\frac{p_E^2 -\frac{k_E^4}{p_E^2} }{(p_E -k_E)^2} \Biggr]
\ \Biggr\}
,
\\
S_2(p_E^2)
&=&
\frac{C_2 (N_c)}{3\pi^2 p_E^2 }
\int_0^\Lambda \hspace{-0.7em} k_E^3 dk_E \int_0^\pi \hspace{-0.5em} \sin^2 \theta d\theta
\frac{\alpha_s [(p_E -k_E)^2]}{\left[ k_E^2 +\Sigma^2 (k_E^2) \right]^2} Z^2 (k_E^2)
\cdot
\left[
2 -\frac{5}{2} \frac{p_E^2 +k_E^2}{(p_E -k_E)^2} +\frac{1}{2} \frac{(p_E^2 -k_E^2)^2}{(p_E -k_E)^4}
\right]
\nonumber\\
&& \hspace{21em} \times
\Biggl\{
\Sigma (k_E^2) S_1 (k_E^2) 
-\left[ k_E^2 - \Sigma^2 (k_E^2) \right] S_2 (k_E^2)
\Biggr\}
,
\\
S_3(p_E^2)
&=&
\frac{C_2 (N_c)}{3\pi^2 p_E^2 }
\int_0^\Lambda \hspace{-0.7em} k_E^3 dk_E \int_0^\pi \hspace{-0.5em} \sin^2 \theta d\theta
\frac{\alpha_s [(p_E -k_E)^2]}{\left[ k_E^2 +\Sigma^2 (k_E^2) \right]^2} Z^2 (k_E^2)
\cdot
\left[
1 + \frac{p_E^2 -2k_E^2}{(p_E -k_E)^2} + \frac{(p_E^2 -k_E^2)^2}{(p_E -k_E)^4}
\right]
\nonumber\\
&& \hspace{10em} \times
\Biggl\{
2 \frac{\Sigma^2 (k_E^2)}{p_E^2} S_1 (k_E^2) 
-4\Sigma (k_E^2) \frac{k_E^2}{p_E^2} S_2 (k_E^2)
-\left[ k_E^2 + \Sigma^2 (k_E^2) \right] \frac{k_E^2}{p_E^2} S_3 (k_E^2)
\Biggr\}
,
\end{eqnarray}
where $(p_E - k_E )^2 = p_E^2 + k_E^2 - 2 p_E k_E \cos \theta$.

By using the Higashijima-Miransky approximation (\ref{eq:Higashijima-Miransky}), it is possible to erase the angular dependence of the running strong coupling $\alpha_s [(p_E - k_E)^2]$, so that the angular integration of Eqs. (\ref{eq:sdes1'}), (\ref{eq:sdes2'}), and (\ref{eq:sdes3'}) can be performed analytically.
We thus obtain
\begin{eqnarray}
S_1 (p_E^2)
&\approx &
1+
\frac{C_2(N_c)}{2\pi}
\int_{p_E}^\Lambda \hspace{-0.7em} k_E dk_E
\frac{\alpha_s [{\rm max}(p_E^2,k_E^2)]}{\left[ k_E^2 +\Sigma^2 (k_E^2) \right]^2} 
(p_E^2 - k_E^2)
\Biggl\{
S_1(k_E^2)
+2\Sigma (k_E^2) S_2(k_E^2)
-\frac{1}{2} \left[ k_E^2 + \Sigma^2 (k_E^2) \right] S_3 (k_E^2)
\Biggr\}
\nonumber\\
&& \hspace{1em}
+
\frac{C_2(N_c)}{2\pi}
\int_0^{p_E} \hspace{-1em} k_E dk_E
\frac{\alpha_s [{\rm max}(p_E^2,k_E^2)]}{\left[ k_E^2 +\Sigma^2 (k_E^2) \right]^2} 
(p_E^2 - k_E^2)
\frac{k_E^4}{p_E^4}
\nonumber\\
&& \hspace{10em} \times
\Biggl\{
\frac{\Sigma^2 (k_E^2)}{k_E^2} S_1(k_E^2)
-2\Sigma (k_E^2) S_2(k_E^2)
-\frac{1}{2} \left[ k_E^2 + \Sigma^2 (k_E^2) \right] S_3 (k_E^2)
\Biggr\}
,
\\
S_2(p_E^2)
&\approx &
\frac{C_2(N_c)}{2\pi}
\int_{p_E}^\Lambda \hspace{-0.7em} k_E dk_E
\frac{\alpha_s [{\rm max}(p_E^2,k_E^2)]}{\left[ k_E^2 +\Sigma^2 (k_E^2) \right]^2} 
\Biggl\{
-\Sigma (k_E^2) S_1 (k_E^2) 
+\left[ k_E^2 - \Sigma^2 (k_E^2) \right] S_2 (k_E^2)
\Biggr\}
\nonumber\\
&&+
\frac{C_2(N_c)}{2\pi}
\int_0^{p_E} \hspace{-1em} k_E dk_E
\frac{\alpha_s [{\rm max}(p_E^2,k_E^2)]}{\left[ k_E^2 +\Sigma^2 (k_E^2) \right]^2} 
\cdot\frac{k_E^4}{p_E^4}
\Biggl\{
-\Sigma (k_E^2) S_1 (k_E^2) 
+\left[ k_E^2 - \Sigma^2 (k_E^2) \right] S_2 (k_E^2)
\Biggr\}
\, , \ \ 
\\
S_3 (p_E^2)
&\approx &
\frac{C_2(N_c)}{2\pi}
\int_0^{p_E} \hspace{-1em} k_E dk_E
\frac{\alpha_s [{\rm max}(p_E^2,k_E^2)]}{\left[ k_E^2 +\Sigma^2 (k_E^2) \right]^2} 
(p_E^2 - k_E^2)
\frac{k_E^4}{p_E^6} 
\nonumber\\
&& \hspace{10em} \times
 \Biggl\{
2\frac{\Sigma^2 (k_E^2)}{k_E^2} S_1 (k_E^2) 
-4\Sigma (k_E^2 ) S_2 (k_E^2)
-\left[ k_E^2 +\Sigma^2 (k_E^2) \right] S_3 (k_E^2) 
\Biggr\}
\, .
\end{eqnarray}
To integrate the angular variable $\theta$, we have used the following formulae:
\begin{eqnarray}
\int_0^\pi \frac{d\theta \, \sin^2 \theta}{a+b\cos \theta} 
&=&
\frac{\pi a}{b^2} \left\{ 1-\sqrt{1-\frac{b^2}{a^2}} \right\}
\, .
\nonumber\\
\int_0^\pi \frac{d\theta \, \sin^2 \theta}{(a+b\cos \theta)^2} 
&=&
\frac{\pi }{a^2} \frac{1}{ \sqrt{1-\frac{b^2}{a^2}} }
-\frac{\pi }{b^2} \left[ 1-\sqrt{1-\frac{b^2}{a^2}} \,\right]
\, .
\end{eqnarray}
We therefore obtain Eqs. (\ref{eq:sdehm1}), (\ref{eq:sdehm2}), and (\ref{eq:sdehm3}).

\twocolumngrid

\end{document}